%% file: main.tex
\documentclass[conference]{IEEEtran}

\usepackage{cite}
\usepackage{dblfloatfix}
\usepackage{amsmath,amssymb,amsfonts}
\usepackage{algorithm}
\usepackage{algpseudocode}
\usepackage{graphicx}
\usepackage[caption=false,font=footnotesize]{subfig}
\usepackage{url}
\usepackage[hidelinks]{hyperref}
\usepackage{microtype}
\usepackage{ragged2e}

\newcommand{\select}{\operatorname{select}}

\begin{document}

\title{PLB: Priority-Aware Load Balancing for Replicated Databases under Constrained Resources}

\author{
\IEEEauthorblockN{
\begin{tabular}{@{}ccc@{}}
\parbox[t]{0.31\textwidth}{\centering Belkis Djeffal} &
\parbox[t]{0.31\textwidth}{\centering Pierre Bourhis} &
\parbox[t]{0.31\textwidth}{\centering Romain Rouvoy}
\end{tabular}
}

\IEEEauthorblockA{
\begin{tabular}{@{}ccc@{}}
\parbox[t]{0.31\textwidth}{
\centering
Inria, Univ. Lille, CNRS\\
UMR 9189 CRIStAL, France\\
belkis.djeffal@inria.fr
}
&
\parbox[t]{0.31\textwidth}{
\centering
CNRS, Univ. Lille, Inria\\
UMR 9189 CRIStAL, France\\
pierre.bourhis@inria.fr
}
&
\parbox[t]{0.31\textwidth}{
\centering
Univ. Lille, Inria, CNRS\\
UMR 9189 CRIStAL, France\\
romain.rouvoy@inria.fr
}
\end{tabular}
}
}

\maketitle

\begin{abstract}
Priority-differentiated services are a standard way for applications to expose different performance expectations, yet database systems still tend to treat all sessions alike. 
When database capacity is fixed, such that replicas cannot be added on demand, and the workload enters contention, this mismatch forces a choice between over-provisioning capacity and allowing lower-priority users to suffer much larger slowdowns. 
In such settings, we propose to enforce priority by controlling how client sessions are assigned to database replicas.

We present \textsf{PLB}, a priority-aware load balancer implemented as a JDBC driver that uses replica assignment to enforce priority differentiation under fixed resources. 
\textsf{PLB} partitions replicas by user group (premium vs.\ freemium) and uses load-based borrowing so that higher-priority users can draw on idle capacity while lower-priority degradation remains controlled. 
We evaluate \textsf{PLB} on a replicated, read-only cluster under OLAP workloads. 
Compared with static dedicated per-priority partitions, \textsf{PLB} sustains utilization above \(\approx 70\%\) in settings where fixed partitions can drive cluster-wide CPU utilization down to \(\approx 35\%\), while keeping latencies close to the best dedicated allocation. Compared with a fully shared round-robin pool, \textsf{PLB} reduces high-priority median latency by about \(12\%\) on average (up to \(28\%\)), while keeping the low-priority median overhead around \(11\%\) and leaving overall median latency close to round-robin.
\end{abstract}

\input{sections/introduction}

\input{sections/problem}
\input{sections/approach}
\input{sections/experimental_setup}
\input{sections/results}

\input{sections/related_work}
\input{sections/threats}

\input{sections/conclusion}

\section*{Acknowledgment}
This work received funding from the France 2030 program, managed by the French
National Research Agency under grant agreement No.~ANR-23-PECL-0003.

\clearpage

\bibliographystyle{IEEEtran}
\bibliography{biblio}

\end{document}

%% file: sections/introduction.tex
\section{Introduction}
Cloud applications implement tiered service plans that associate different user classes with distinct performance expectations and economic value~\cite{smartsla,activesla}.
This differentiation is reflected in various quality-of-service
(QoS) dimensions---including features, response latency, and
resolution---and is explicitly enforced at the application layer.
However, at the database layer, this distinction is not preserved: database systems optimize queries without directly encoding the application-level service tier of the user.
This mismatch becomes particularly significant under resource contention. 
When database capacity is abundant, uniform optimization may suffice; all users can experience good performance regardless of priority.
Under contention, however, the system must preserve better latency for high-priority users while allowing lower-priority users to experience predictable, graceful degradation.

This paper focuses on read-only database deployments operating under a fixed replica budget. 
Such deployments can arise in analytical and machine-learning platforms, where a shared feature store or read-only database may simultaneously serve latency-sensitive online inference and more delay-tolerant offline training or evaluation over the same data. 
Although the workload is read-only, the services sharing these replicas may still differ substantially in their latency requirements. 
Adding resources through elastic scaling might appear to address this problem, but in many settings it is impractical: provisioning a new replica introduces delay due to provisioning and data replication, while elastic scaling itself incurs operational complexity and cost~\cite{pstore,edgedb}.
Moreover, on-premise deployments, edge environments, and resource-constrained cloud offerings may not permit capacity to be added on demand.

Prior work has addressed priority through admission control~\cite{smartsla,activesla,bouncer} and database engine scheduling~\cite{selftuning,dbmspredict,lsched}. 
However, widely deployed systems such as PostgreSQL delegate CPU scheduling to the operating system and do not expose native mechanisms to assign query priorities or isolate priority classes within the engine~\cite{PostgreSQLWiki_Priorities_2017}. 
In replicated databases, the load balancer mediates access to replicas and therefore provides a natural locus for priority-aware routing without requiring changes to the database engine~\cite{Designing,replication_vldb}. 
At that point, the problem is no longer whether priority should be enforced, but how replica capacity should be allocated across classes so that sharing does not erase isolation. 
In practice, two deployment choices are common. 
Dedicated clusters provide strong isolation, but waste capacity when demand is imbalanced: replicas reserved for one class may remain underutilized while the other class saturates. 
At the opposite extreme, a shared cluster achieves high utilization by allowing all sessions to access all replicas, but without priority-aware routing, all sessions compete equally for capacity, exposing high-priority traffic to the same slowdowns as best-effort traffic. 
Neither choice is satisfactory: the former sacrifices efficiency, while the latter sacrifices service differentiation.

To address this gap, we introduce \textsf{PLB}, a priority-aware session assignment strategy for replicated databases under fixed resource constraints. The key idea is to preserve priority differentiation while dynamically sharing a fixed replica budget as the workload mix shifts. Concretely, \textsf{PLB} organizes replicas into three groups: \textsf{Premium}, \textsf{Freemium}, and \textsf{Mixed}, the latter acting as a dynamic buffer between the first two. When high-priority demand approaches saturation, \textsf{PLB} can temporarily reassign underutilized capacity; when pressure recedes, borrowed replicas are returned so that capacity sharing does not erase long-term isolation. \textsf{PLB} is implemented as a JDBC middleware layer that operates at session admission time, aligning with standard connection-level routing practice and avoiding modifications to the database engine internals~\cite{PgpoolII_website,HAProxy_website,pgpool_connection_pooling_2025}.

To evaluate this approach, we use the TPC-H decision-support benchmark~\cite{tpch2022} on a replicated, read-only PostgreSQL cluster under varying load levels and Premium:Freemium client mixes. 
We compare \textsf{PLB} against two common deployments: static dedicated clusters and a shared priority-agnostic round-robin pool (\textsf{RR}). 
Our results show that \textsf{PLB} reacts better to changes in workload composition than fixed partitions: when the live Premium:Freemium mix diverges from the partition for which the deployment was sized, dedicated clusters can strand a large fraction of capacity and drive cluster-wide CPU utilization down to around 35\%, whereas \textsf{PLB} keeps utilization above about 70\% while maintaining latencies close to the best dedicated allocation. 
At the same time, \textsf{PLB} enforces service differentiation without substantially increasing overall cost: in our experiments, it improves \textsf{Premium} median latency by roughly 12\% on average (up to 28\%) and \textsf{Premium} $p95$ latency by about 10\% (up to 14\%), while keeping the \textsf{Freemium} penalty around 10\% and leaving overall latency close to \textsf{RR}.

This paper delivers four contributions:
(1) We formalize the isolation-efficiency trade-off in priority-aware replica allocation for replicated databases. 
(2) We present \textsf{PLB}, a priority-aware session assignment policy with dynamic borrowing that achieves high utilization and service differentiation simultaneously. 
(3) We provide an open-source JDBC middleware implementation, and
(4) we experimentally validate PLB across different workload configurations.
The remainder of this paper is organized as follows. 
Sections~2--3 formalize the problem and introduce the \textsf{PLB} design. 
Sections~4--5 cover the experimental setup and empirical evaluation. 
Section~6 surveys related work, Section~7 discusses threats to validity, and Section~8 concludes.

%% file: sections/problem.tex
\section{Problem Statement}\label{sec:problem}

\subsection{System \& Workload Model}\label{sec:sysmodel}
We consider a cluster of $K$ read-only replicas of a database, each running on a separate node. 
All replicas store the same data and share the same DBMS configuration; they only differ in their current load. 
The number of replicas $K$ and their hardware configuration are fixed for the duration of the deployment and cannot be scaled on demand.

Clients interact with the system through \emph{sessions}---i.e., sequences of queries issued over persistent database connections. 

Each session is tagged with a priority class, \textsf{Premium} or \textsf{Freemium}, reflecting the client's application-level service plan. 

Once a session is assigned to a replica, all its queries are executed on that replica; we do not migrate sessions between replicas, nor do we modify DBMS internals, such as the query optimizer or executor. 

Let $\Sigma(t)$ denote the system state observed by the routing layer at time $t$ (e.g., per-replica and per-class active-session counts).
A routing strategy $\Gamma$ assigns each incoming session to one replica in $R=\{1,\dots,K\}$ as a function of the session's priority class and $\Sigma(t)$. 
Formally, when a session $s$ arrives at time $t$, the strategy chooses a replica $\Gamma\!\bigl(\mathsf{class}(s),\Sigma(t)\bigr)\in R$ that will serve that session.

\subsection{Capacity, Load \& Contention}
We model the current load on a replica as the number of active sessions. 

Let $\text{load}_P(r,t)$ and $\text{load}_F(r,t)$ denote the numbers of active \textsf{Premium} and \textsf{Freemium} sessions assigned to replica $r \in R$ and not yet completed at time $t$. 

We write
\[
  \text{load}(r,t) = \text{load}_P(r,t) + \text{load}_F(r,t)
\]
for the total number of active sessions on $r$. 

In our implementation, these per-class session counts are the primary load signals available to the routing layer.

For each class $c \in \{P,F\}$, we configure a per-replica saturation threshold $\theta_c$.
We say that replica $r$ has \emph{headroom} for class $c$ at time $t$ when $\text{load}_c(r,t) < \theta_c$, and that it is \emph{saturated for class $c$} otherwise. 
These empirically calibrated thresholds, described in Section~\ref{sec:experimental-setup}, are used as control signals by the routing policy.

We refer to regimes in which many replicas approach or exceed their per-class thresholds as \emph{contention regimes}. 
In such regimes, the routing strategy must enforce differentiation between \textsf{Premium} and \textsf{Freemium} while coping with fixed capacity. A practical solution must therefore preserve priority differentiation while keeping the fixed replica budget well utilized as the Premium:Freemium demand mix changes.

\subsection{Key Performance Metrics}\label{sec:metrics}
Because we consider read-intensive workloads (e.g., OLAP), we evaluate performance primarily in terms of query latency.
For each workload run, we collect latency distributions separately for \textsf{Premium}, \textsf{Freemium}, and all queries, denoted respectively by $\psi_{\textsf{P}}$, $\psi_{\textsf{F}}$, and $\psi_{\textsf{All}}$. 
In the evaluation, we summarize these distributions using central and tail statistics, depending on the experiment, to compare both typical and degraded behavior under contention.

\subsection{Service Level Objectives}\label{sec:slo}
We characterize the behavior sought from a routing strategy under fixed resources through three objectives.

\hypertarget{obj:o1}{}
\paragraph*{O1: Prioritized Premium latency under contention}
When the cluster enters a contention regime, \textsf{Premium} queries should systematically experience shorter latency than \textsf{Freemium} queries. 
As load increases and replicas saturate, $\psi_{\textsf{P}}$ may grow, but it should grow more slowly than $\psi_{\textsf{F}}$, so that \textsf{Premium} sessions remain closer to their expected QoS.

\hypertarget{obj:o2}{}
\paragraph*{O2: Controlled Freemium degradation}
While \textsf{Freemium} queries may be deprioritized under heavy load, they should not be starved. 
Compared to priority-agnostic strategies, such as round-robin, $\psi_{\textsf{F}}$ under our design may be higher, but it should degrade smoothly rather than through pathological spikes. 
In the evaluation, we assess this objective empirically by comparing $\psi_{\textsf{F}}$ against priority-agnostic baselines across various load levels.

\hypertarget{obj:o3}{}
\paragraph*{O3: Efficient sharing under fixed capacity}
Under a fixed replica budget, the routing strategy should sustain high utilization of the replica pool while keeping load well balanced across replicas as the Premium:Freemium demand mix varies. This objective captures the ability to adapt to changing demand without leaving part of the replica budget underused.

\subsection{Problem Formulation}

We formulate the addressed problem as:
Given a fixed pool of $K$ read-only replicas, a stream of incoming sessions tagged as \textsf{Premium} or \textsf{Freemium}, and per-replica load observations based on active sessions, we seek a session-level routing strategy $\Gamma$ that satisfies \hyperlink{obj:o1}{O1}--\hyperlink{obj:o3}{O3}: it should preserve prioritized latency for \textsf{Premium} sessions under contention, keep \textsf{Freemium} degradation controlled relative to priority-agnostic baselines, and reuse idle replicas across classes without elastic scaling. 

%% file: sections/approach.tex
\section{Priority-Aware Load Balancing}\label{sec:approach}

We consider the model defined in Section~\ref{sec:sysmodel}: a fixed pool of read-only replicas and session-level routing for two priority classes (\textsf{Premium}, \textsf{Freemium}). 
We operate at session admission time, which matches common connection-level routing practice in replicated database middleware and avoids per-query routing overhead. 
Our goal is to enforce priority-aware routing under fixed resources by combining class separation with temporary capacity sharing under pressure.

\subsection{High-Level Overview}

\begin{figure*}[t]
  \centering
  \subfloat[Routing dynamics over time.]{
    \includegraphics[width=0.62\textwidth]{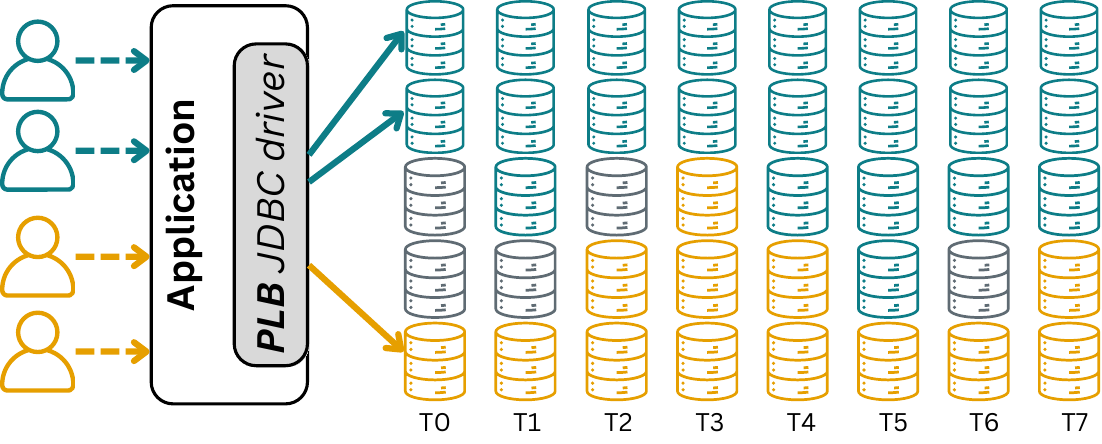}
    \label{fig:plb-dynamics}
  }
  \hfill
  \subfloat[Conceptual role transitions.]{
    \includegraphics[width=0.28\textwidth]{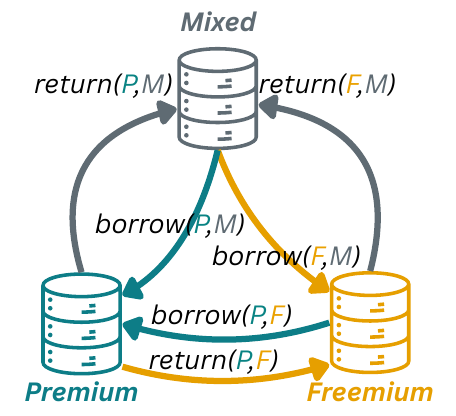}
    \label{fig:plb-transitions}
  }
  \caption{Overview of \textsf{PLB}. (a) Routing dynamics over time. (b) Conceptual borrow/return transitions between \emph{Premium}, \emph{Freemium}, and \emph{Mixed} roles.}
  \label{fig:plb-overview}
\end{figure*}

\textsf{PLB} partitions replicas into three pools: a \emph{Premium pool}, a \emph{Freemium pool}, and a \emph{Mixed pool}. 
Sessions are first routed to the pool currently assigned to their class as long as that pool still has headroom. 
When a class exhausts its headroom, the router may temporarily borrow capacity from the \emph{Mixed pool} or, under stricter conditions, from the opposite dedicated pool. 
Figure~\ref{fig:plb-overview} summarizes the design. 
Figure~\ref{fig:plb-dynamics} gives an example of how replica assignments evolve over time, while Figure~\ref{fig:plb-transitions} summarizes the possible role transitions; the admission and return conditions are described below.

The policy is asymmetric across classes. 
\textsf{Premium} may borrow from the \emph{Mixed pool} and, when a minimum reserve for \textsf{Freemium} is preserved, from the \emph{Freemium pool}. 
\textsf{Freemium} may also borrow from the \emph{Mixed pool}, but borrowing from the \emph{Premium pool} is permitted only when the donor replica is empty of \textsf{Premium} sessions and the Premium pool target remains protected.

\subsection{Replica Roles \& Pool Partitioning}

At any time, the router maintains three replica sets: the
\emph{Premium pool}, the \emph{Freemium pool}, and the
\emph{Mixed pool}, denoted by \(\mathcal{R}_P\),
\(\mathcal{R}_F\), and \(\mathcal{R}_M\), respectively. 
The total replica budget is fixed, with
\(K = |\mathcal{R}_P| + |\mathcal{R}_F| + |\mathcal{R}_M|\).
The initial split across the three pools is configured at startup.

Replicas in \(\mathcal{R}_P\) serve \textsf{Premium} sessions by
default, replicas in \(\mathcal{R}_F\) serve \textsf{Freemium}
sessions by default, and replicas in \(\mathcal{R}_M\) form a
flexible buffer that can donate capacity in either direction.
Borrowing is implemented as a temporary role change: the donor
replica is moved into the borrower's pool and is counted there until
it is released. Returns are also implemented as role changes: an
eligible borrowed replica is moved either to the pool whose target is
not yet satisfied or to the \emph{Mixed pool}.

\subsection{Load Metrics \& Control Signals}

Routing decisions are taken at session admission using per-replica,
per-class active-session counts. For a replica \(r\), we write
\(\text{load}_P(r)\) and \(\text{load}_F(r)\) for the numbers of active
\textsf{Premium} and \textsf{Freemium} sessions on \(r\), observed at
admission time. We also write
\[
\text{load}(r)=\text{load}_P(r)+\text{load}_F(r)
\]
for the total number of active sessions on \(r\).

The routing policy uses four control signals:
\begin{itemize}
  \item \textbf{Saturation thresholds} \(\theta_P,\theta_F\): thresholds
  used to decide when a class has exhausted the headroom of its
  current pool and may attempt borrowing.

  \item \textbf{Return watermarks} \(\tau_c = \lfloor \alpha_c \theta_c \rfloor\):
  lower thresholds used to decide when pressure has eased enough
  to release borrowed capacity.

  \item \textbf{Freemium floor} \(\kappa_F\): minimum number of replicas
  kept in the \textsf{Freemium} pool when \textsf{Premium} borrows
  from that pool.

  \item \textbf{Pool targets} \(K_P,K_F\): configured target sizes used when deciding whether a returned replica should go to a class pool or to the \emph{Mixed pool}.
\end{itemize}

\subsection{Admission and Return Policy}

\paragraph*{Common principles}
We call a replica \emph{\textsf{Premium}-empty} when \(\text{load}_P(r)=0\). 
The following rules apply to both priority
classes.
First, a class may borrow only when all active replicas currently
assigned to that class are at or above the saturation threshold. Second, role changes are applied only at session admission time: whether a replica is borrowed or released, existing sessions are not moved, interrupted, or killed; only future session assignments see the new role.
Third, a borrowed replica is released only after pressure on the borrowing class has dropped below the return watermark and after the replica no longer hosts sessions of that borrowing class.

In the admission rules below, when PLB selects among replicas
already assigned to class \(c\), it uses \(\text{load}_c(r)\). For replicas
currently borrowed for class \(c\), it uses the total load
\(\text{load}(r)\), so that residual sessions from the previous role are
counted.

\paragraph*{\textsf{Premium} admission}
A \textsf{Premium} session is first routed to an eligible replica in
\(\mathcal{R}_P\), choosing the least-loaded replica whose
\textsf{Premium} load is below \(\theta_P\). If no such replica exists,
the router attempts to borrow from the \emph{Mixed pool}. Among
Mixed replicas, PLB first prefers replicas that do not already host
\textsf{Premium} sessions, so that Premium traffic is spread over
additional replicas; it then chooses the one with the smallest
\textsf{Freemium} load.

If the Mixed pool is empty, PLB may borrow from the
\emph{Freemium pool}. 
This is allowed only if the Freemium floor is preserved, i.e.,
\(|\mathcal{R}_F|-1\ge\kappa_F\), and if the candidate Freemium
replica is less loaded than the best current Premium choice under
the load convention above.

Among eligible Freemium donors, PLB prefers replicas below the
Freemium return watermark and then chooses the least-loaded one.
If no donor is eligible, the session is routed to the least-loaded
replica in \(\mathcal{R}_P\).

\paragraph*{\textsf{Freemium} admission}
A \textsf{Freemium} session is first routed to a replica in
\(\mathcal{R}_F\) whose \textsf{Freemium} load is below
\(\theta_F\), choosing the least-loaded eligible replica. If no such
replica exists, the router first considers donors in the
\emph{Mixed pool}, choosing the replica with the fewest active
\textsf{Premium} sessions and then the smallest active
\textsf{Freemium} load. If Freemium still has no eligible replica and
the Mixed pool is empty, the router may borrow from the
\emph{Premium pool}, but only from replicas that are
\textsf{Premium}-empty and only if the Premium pool target is
preserved, i.e., \(|\mathcal{R}_P|-1\ge K_P\).
If no donor is eligible, the session is routed to the least-loaded
replica in \(\mathcal{R}_F\).

\paragraph*{Return}
A borrowed replica is returned only when two conditions hold:
pressure on the borrowing class has eased, and the borrowed replica
no longer hosts sessions of the borrowing class. Pressure is
considered eased when some active replica assigned to class \(c\)
falls below the return watermark \(\tau_c\).

When a borrowed replica is returned, it normally goes to the
\emph{Mixed pool}. The exception is when a class pool is below its
configured target: a \texttt{borrowed-to-Premium} replica is assigned
back to the \emph{Freemium pool} if \(|\mathcal{R}_F|<K_F\), and to
the \emph{Mixed pool} otherwise. Symmetrically, a
\texttt{borrowed-to-Freemium} replica is assigned back to the
\emph{Premium pool} if \(|\mathcal{R}_P|<K_P\), and to the
\emph{Mixed pool} otherwise.

Algorithm~\ref{alg:plb-assignment} summarizes the session-admission path. 
The functions \(\select_{M\to P}\), \(\select_{F\to P}\), and
\(\select_{M\to F}\) use the donor-ordering rules described above.

\begin{algorithm}[t]
\caption{Priority-aware session admission}
\label{alg:plb-assignment}
\small
\RaggedRight
\begin{algorithmic}[1]
\Function{assignSessionToReplica}{$p$}

  \If{$p=\texttt{premium}$}
    \State \(D \gets\) replicas in \(\mathcal{R}_P\) below \(\theta_P\)
    \If{\(D\neq\emptyset\)}
      \State \Return least-loaded replica in \(D\)
    \EndIf

    \If{\(\mathcal{R}_M\neq\emptyset\)}
      \State \(r_b \gets \select_{M\to P}(\mathcal{R}_M)\)
      \State change role of \(r_b\) to \texttt{borrowed-to-Premium}
      \State \Return \(r_b\)
    \EndIf

    \If{\(|\mathcal{R}_F|-1\ge\kappa_F\)}
      \State \(m_P \gets\) load of the least-loaded replica in \(\mathcal{R}_P\)
      \State \(C_F \gets\) Freemium donors with load below \(m_P\)
      \If{\(C_F\neq\emptyset\)}
        \State \(r_b \gets \select_{F\to P}(C_F)\)
        \State change role of \(r_b\) to \texttt{borrowed-to-Premium}
        \State \Return \(r_b\)
      \EndIf
    \EndIf

    \State \Return least-loaded replica in \(\mathcal{R}_P\)

  \Else
    \Comment{\textsf{Freemium}}
    \State \(D \gets\) replicas in \(\mathcal{R}_F\) below \(\theta_F\)
    \If{\(D\neq\emptyset\)}
      \State \Return least-loaded replica in \(D\)
    \EndIf

    \If{\(\mathcal{R}_M\neq\emptyset\)}
      \State \(r_b \gets \select_{M\to F}(\mathcal{R}_M)\)
      \State change role of \(r_b\) to \texttt{borrowed-to-Freemium}
      \State \Return \(r_b\)
    \EndIf

    \State \(C_P \gets \{r\in\mathcal{R}_P : \text{load}_P(r)=0\}\)
    \If{\(C_P\neq\emptyset\) \textbf{and} \(|\mathcal{R}_P|-1\ge K_P\)}
      \State \(r_b \gets\) least-loaded replica in \(C_P\)
      \State change role of \(r_b\) to \texttt{borrowed-to-Freemium}
      \State \Return \(r_b\)
    \EndIf

    \State \Return least-loaded replica in \(\mathcal{R}_F\)
  \EndIf

\EndFunction
\end{algorithmic}
\end{algorithm}
\normalsize
\justifying

\subsection{Implementation \& Complexity}

\paragraph*{Configuration}
The policy is instantiated through the initial pool split and the control parameters introduced above. 
In practice, the main controls are the per-class saturation thresholds \(\theta_c\), which determine when borrowing is triggered; the return watermark \(\tau_c\) is derived from \(\theta_c\), while \(\kappa_F\) and the pool targets \(K_P,K_F\) govern how conservatively capacity is shared and returned. 
For the experiments, these values are calibrated once per replica budget and then kept fixed across the full evaluation grid, as described in Section~\ref{sec:experimental-setup}.

\paragraph*{Implementation}
\textsf{PLB} is implemented as a JDBC middleware that intercepts connection establishment. 
Replicas and the initial pool split are provided through configuration files, while role changes are maintained internally by the router.
At session admission, the middleware identifies the session priority, reads the current per-replica state (role and per-class session counts), and applies Algorithm~\ref{alg:plb-assignment}. 
Per-replica counters are updated atomically when a connection opens or closes. 
The design is compatible with standard JDBC connection pools, since sessions remain pinned to their assigned replica for their lifetime.

\paragraph*{Complexity}
Each admission scans at most \(K\) replicas. 
Routing therefore runs in \(O(K)\) time, while the middleware maintains \(O(K)\) state.

%% file: sections/experimental_setup.tex
\section{Experimental Setup}\label{sec:experimental-setup}

To assess the effectiveness of our priority-aware load-balancing strategy under realistic conditions, we deploy it on a physically distributed PostgreSQL cluster configured as read-only replicas and simulate session priority and contention using a custom benchmarking framework.

\subsection{Infrastructure \& Workload}

Experiments are conducted on Grid'5000~\cite{grid5000}, using homogeneous physical machines from the \texttt{gros} cluster at the Nancy site. 
Each node is equipped with an 18-core Intel Xeon Gold 5220 CPU and 96~GiB of RAM. 
Replicas are provisioned as identical clones of the same dataset, while the workload generator and JDBC middleware run on a dedicated injector node in order to isolate benchmarking overhead from database execution.

All nodes run Ubuntu 22.04~LTS, PostgreSQL~14, and OpenJDK~21. 
PostgreSQL is tuned once for read-heavy workloads, starting from a \texttt{PGTune}~\cite{pgtuneX} configuration and adjusting key settings such as shared buffers and parallelism limits through workload-driven tests.

Our workload is based on TPC-H~\cite{tpch2022} at a scale factor of 10, representing read-only decision-support queries. 
We use \textsc{BenchBase}~\cite{DifallahPCC13}, extended to support fixed session priorities (\texttt{premium} or \texttt{freemium}), Poisson arrivals, and configurable session lifetimes. 
Within each session, queries are selected uniformly at random from the TPC-H query set and executed sequentially.

In our modified \textsc{BenchBase}, a \emph{terminal} corresponds to one JDBC session pinned to a single replica for its lifetime. 
For a target of \(T\) terminals over duration \(D\), arrivals follow an exponential process of rate \(\lambda = T/D\), and each terminal lives for a duration uniformly sampled in \([10,30]\) seconds. 
Because sessions have finite lifetimes and include think time, the instantaneous number of concurrently executing PostgreSQL backends per replica remains much smaller than \(T\); for example, with five replicas and \(T{=}1000\), we observe a median of about 50 active backends per replica.

\subsection{\textsf{PLB} Configuration}\label{subsec:plb-config}

In all experiments, we use a single \textsf{PLB} configuration per replica budget. 
This configuration is chosen once on the same TPC-H SF10 workload as in our evaluation, using a subset of client counts and Premium$:$Freemium mixes that cover both moderate and high load. 
Once chosen, the saturation thresholds and other \textsf{PLB} parameters are kept fixed for the full experimental grid.

\paragraph{\textbf{Saturation and return parameters}}
The per-class saturation thresholds \(\theta_P\) and \(\theta_F\) determine when a replica is considered to have exhausted headroom for a given class. 
We select these thresholds by sweeping \(\theta\) over a small range on the calibration subset and measuring end-to-end latency for both priority classes. 
Very small values trigger borrowing too early and lead to unnecessary churn, whereas very large values delay borrowing and leave the mixed pool underused. 
We therefore set \(\theta_P=\theta_F=10\) for all reported experiments. 
Because these thresholds only determine when borrowing is triggered, we keep them symmetric across classes and express priority through pool partitioning, borrowing rules, and asymmetric return watermarks.

Return watermarks \(\tau_c=\lfloor \alpha_c \theta_c \rfloor\) and pool targets \(K_P,K_F\) determine when borrowed replicas can be returned and how the pool partition is restored once pressure has eased. 
We fix \(\alpha_P=0.60\) and \(\alpha_F=0.80\), so that replicas borrowed by \textsf{Premium} become eligible to return at a lower load than replicas borrowed by \textsf{Freemium}. 
For the five-replica budget, we initialize the pools as \((|\mathcal{R}_P|,|\mathcal{R}_F|,|\mathcal{R}_M|)=(2,2,1)\), set \(K_P=K_F=2\), and keep a small Freemium floor \(\kappa_F=1\) to avoid giving away the last Freemium replica during borrowing.

\subsection{Evaluation Protocol}\label{subsec:evaluation-space}

We evaluate \textsf{PLB} over a grid of load levels and priority mixes, and against the baselines introduced in Section~\ref{subsec:results-baselines}. 
The number of terminals \(T\) ranges from \(30\) to \(2{,}500\), and for each \(T\) we vary the Premium share from \(20\%\) to \(80\%\) in steps of 10 percentage points, yielding \(161\) distinct \((T,\text{Premium share})\) configurations.

All configurations are executed sequentially in a single batch, with a single warmup before the batch begins. 
Each configuration runs for 3 minutes and is repeated 10 times. 
During each run, we collect per-query latencies, per-session statistics, and per-replica CPU usage. 
Latency distributions for each class and the overall distribution are derived as in Section~\ref{sec:metrics}, and CPU utilization and imbalance are computed from \texttt{pidstat} samples, as detailed in Section~\ref{sec:results}. 
All deployment and benchmarking scripts, including our JDBC middleware and modified \textsc{BenchBase}, are publicly available.\footnote{\url{https://zenodo.org/records/20233910}}

%% file: sections/results.tex
\section{Empirical Results}\label{sec:results}
\subsection{Baselines, Configurations \& Objectives}\label{subsec:results-baselines}
We evaluate \textsf{PLB} under fixed replica budgets and compare it with two deployment choices that a provider can implement today.
For a given budget \(K\), we consider:
\begin{enumerate}
  \item \textbf{Shared, priority-agnostic pool (\(K\)\textsf{RR}).}
        All client sessions (Premium and Freemium) share the same set of \(K\) replicas, and connections are distributed using a standard policy such as round-robin.
        We denote this baseline by \emph{\(K\)\textsf{RR}} (e.g., \emph{\textsf{5RR}} when \(K=5\)).

  \item \textbf{Dedicated per-class partitions (\(r\)\textsf{RR}+\(f\)\textsf{RR}, \(r+f=K\)).}
        The \(K\) replicas are statically partitioned between the two priority classes.
        The left-hand term counts \textsf{Premium} replicas and the right-hand term \textsf{Freemium} replicas; within each subset, sessions are routed by round-robin.
        We denote such a partition by \emph{\(r\)\textsf{RR}+\(f\)\textsf{RR}}---e.g., \emph{\textsf{3RR+2RR}} (three \textsf{Premium}-only replicas and two \textsf{Freemium}-only replicas) or \emph{\textsf{4RR+1RR}} when \(K=5\).
\end{enumerate}

\textsf{PLB} itself uses all \(K\) replicas in a single shared pool but routes sessions with the priority-aware logic of Section~\ref{sec:approach}.
We denote this configuration by \emph{\(K\)PLB} (e.g., \emph{5PLB}, \emph{6PLB}).

Our evaluation focuses on the objectives defined in Section~\ref{sec:problem}: prioritized \textsf{Premium} latency under contention, controlled \textsf{Freemium} degradation, and efficient sharing under fixed capacity.

In a shared \(K\)\textsf{RR} pool, sessions are routed without regard to priority, with the result that \textsf{Premium} and \textsf{Freemium} latencies remain closely coupled. 
Dedicated per-class partitions provide stronger isolation, but require fixing the split \emph{a priori} and become brittle once the \textsf{Premium:Freemium} mix drifts. 
The following subsections are organized around these objectives. We first examine whether \textsf{PLB} reproduces the class-level behavior of dedicated partitions across loads and mixes (\hyperlink{obj:o1}{O1}, \hyperlink{obj:o2}{O2}). We then study whether it uses the fixed replica budget more efficiently as the \textsf{Premium:Freemium} mix changes, while revisiting class-level latency at a fixed heavy load (\hyperlink{obj:o3}{O3}, \hyperlink{obj:o1}{O1}, \hyperlink{obj:o2}{O2}). Finally, we compare \textsf{PLB} with shared round-robin to quantify the \textsf{Premium} gains and the induced cost for \textsf{Freemium} (\hyperlink{obj:o1}{O1}, \hyperlink{obj:o2}{O2}).

\subsection{PLB vs. Dedicated Partitions}\label{subsec:results-plb-vs-splits}

\begin{figure}[t]
  \centering
  \includegraphics[width=0.92\columnwidth]{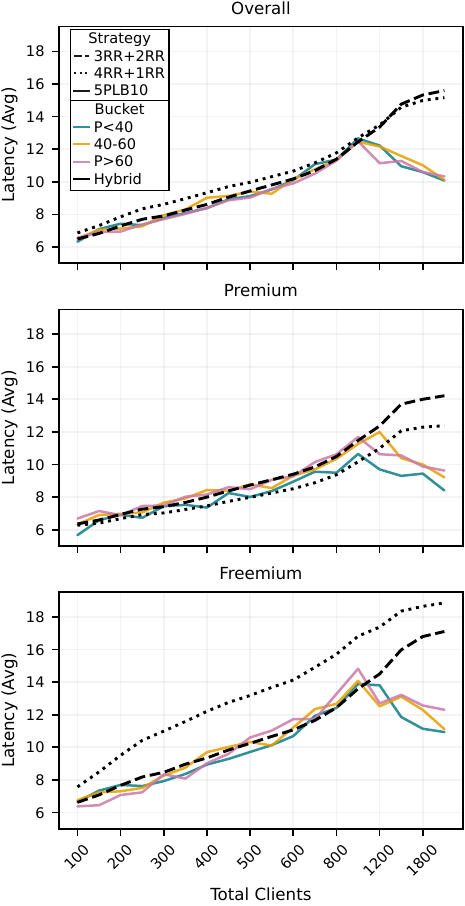}
    \vspace{-6pt}
  \caption{Per-class average latency vs.\ terminals for \textsf{5PLB} and dedicated clusters (three mixes).}
  \label{fig:plb-dedicated-per-mix-avg-col}
\end{figure}

We compare \textsf{5PLB} with two dedicated 5-replica partitions, \textsf{3RR+2RR} and \textsf{4RR+1RR}, over terminal counts from $100$ to $1,800$ and three representative \textsf{Premium:Freemium} mixes. Figure~\ref{fig:plb-dedicated-per-mix-avg-col} reports the corresponding average latencies for Overall, \textsf{Premium}, and \textsf{Freemium} queries. Because these dedicated baselines are hybrid constructions built from separate Premium-only and Freemium-only runs, we report average latency here: combined means can be computed from the component runs, whereas combined medians cannot. We then examine whether \textsf{PLB} matches the class-level behavior of the best static partition across load levels and demand mixes, which speaks directly to \hyperlink{obj:o1}{O1} and \hyperlink{obj:o2}{O2}.

At light load (\(T \lesssim 200\)), all strategies exhibit nearly identical latencies, reflecting that replicas are far from saturation. 
Under contention, their behavior diverges. 
Overall average latency under \textsf{5PLB} remains close to \textsf{3RR+2RR} and clearly below \textsf{4RR+1RR}, indicating that enforcing priority does not incur a visible aggregate cost.   

On the \textsf{Premium} side, \textsf{5PLB} behaves as if \textsf{Premium} were served by three to four dedicated replicas. 
When \textsf{Premium} demand is low, its latency approaches that of \textsf{4RR+1RR}; as \textsf{Premium} traffic grows, it settles near \textsf{3RR+2RR}. 
In practice, \textsf{Premium} latency under \textsf{5PLB} stays close to the three-replica dedicated baseline and often approaches the four-replica configuration when \textsf{Premium} demand is low, which is consistent with \hyperlink{obj:o1}{O1}.

\textsf{Freemium} exhibits the symmetric effect. 
When \textsf{Freemium} dominates, \textsf{5PLB} slightly improves on \textsf{3RR+2RR}, since borrowing allows it to exploit more capacity than a fixed one- or two-replica budget. 
As \textsf{Premium} load increases, \textsf{Freemium} latency under \textsf{5PLB} rises, but remains much closer to the behavior of the dedicated configurations that preserve two replicas for \textsf{Freemium} than to the one-replica worst case. 
This controlled slowdown of the low-priority class, while the high-priority class continues to track the best dedicated configuration, matches \hyperlink{obj:o2}{O2}.

Taken together, these curves show that \textsf{5PLB} satisfies \hyperlink{obj:o1}{O1} and \hyperlink{obj:o2}{O2}: across loads and mixes, \textsf{Premium} follows the best of \textsf{3RR+2RR} and \textsf{4RR+1RR} as contention increases, while \textsf{Freemium} remains close to a dedicated two-replica budget rather than collapsing to a one-replica worst case.

\subsection{PLB vs. Dedicated Partitions at Fixed Load}\label{subsec:results-plb-vs-splits-fixed-load}

We compare \(K\)\textsf{PLB} with dedicated per-class partitions at a fixed heavy load. Specifically, we set the total number of terminals to \(T = 1000\) and vary the \textsf{Premium:Freemium} mix, in order to assess both resource efficiency (\hyperlink{obj:o3}{O3}) and per-class performance relative to static partitions (\hyperlink{obj:o1}{O1}, \hyperlink{obj:o2}{O2}).

CPU metrics are obtained from \texttt{pidstat} samples. 
For each replica and second, we sum the user and system CPU percentages across \texttt{postgres} processes, divide by \(100 \cdot C\) (where \(C\) is the number of cores), and average over time to obtain a per-replica mean CPU fraction \(u_i\). 
For each configuration, cluster-level CPU utilization is \(\overline{u} = \frac{1}{K}\sum_i u_i\), and the coefficient of variation \(\textit{CV}(u) = \sigma(u_i)/\mu(u_i)\) measures load imbalance across replicas. Both metrics are averaged across repeated runs.

Dedicated partitions are derived from single-priority round-robin experiments. 
For each \textsf{Premium:Freemium} mix \((P,F)\), we run \textsf{3RR} and \textsf{4RR} with \(P\) \textsf{Premium} terminals and \textsf{2RR} and \textsf{1RR} with \(F\) \textsf{Freemium} terminals, collect the per-replica means \(u_i\), and use them as empirical pools. We then approximate \textsf{3RR+2RR} and \textsf{4RR+1RR} by bootstrap resampling from the corresponding Premium-only and Freemium-only pools, respectively, and report the average over 400 resamples.

\begin{figure}[ht]
  \centering
  \includegraphics[width=0.9\columnwidth]{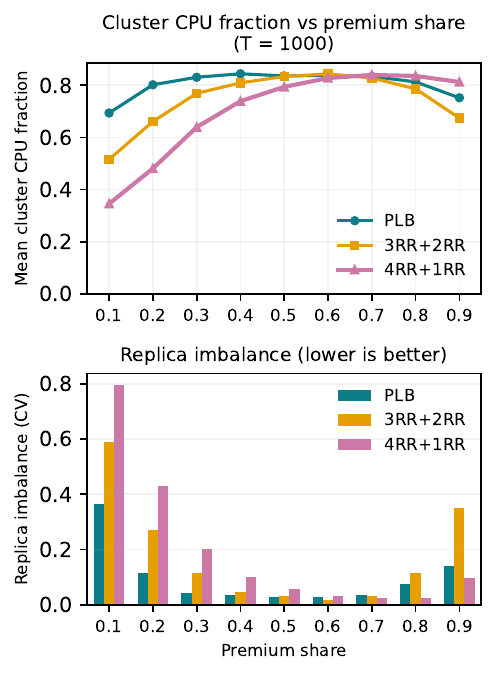}
  \caption{CPU utilization and replica imbalance at \(T = 1000\) terminals for \textsf{5PLB} and dedicated partitions.}
  \label{fig:cpu-util-cv-1000}
\end{figure}

\paragraph*{\textbf{CPU utilization and imbalance}}
Figure~\ref{fig:cpu-util-cv-1000} shows the mean cluster CPU fraction and the coefficient of variation of per-replica CPU across \textsf{Premium} shares at \(T = 1000\). 
With \textsf{5PLB}, mixes between roughly 0.2 and 0.8 \textsf{Premium} share keep the cluster in a high-utilization, low-imbalance regime: \(\overline{u}\) stays between about 0.80 and 0.84, and \(\textit{CV}(u)\) remains below 0.12. 
Even in the most skewed mixes, \textsf{5PLB} does not drop below \(\overline{u} \approx 0.69\), with \(\textit{CV}(u)\) well under 0.4.

In contrast, dedicated partitions are much more brittle. 
When the live \textsf{Premium:Freemium} mix diverges from the configured split, replicas tied to the minority class are underutilized while majority replicas saturate. 
For example, in a Freemium-heavy mix (Premium share 0.1), \textsf{4RR+1RR} falls to about \(\overline{u} \approx 0.35\) with \(\textit{CV}(u) \approx 0.80\), whereas \textsf{5PLB} still reaches \(\overline{u} \approx 0.69\) with much lower imbalance (\(\textit{CV}(u) \approx 0.36\)). 
Symmetrically, when \textsf{Premium} dominates (Premium share 0.9), \textsf{3RR+2RR} leaves part of the cluster underused (\(\overline{u} \approx 0.67\), \(\textit{CV}(u) \approx 0.59\)), whereas \textsf{5PLB} keeps more replicas busy and better balanced (\(\overline{u} \approx 0.75\), \(\textit{CV}(u) \approx 0.14\)). 
At \(T = 1000\), these results support \hyperlink{obj:o3}{O3}: \textsf{5PLB} maintains high, well-balanced CPU utilization across mixes, whereas fixed partitions can strand a large fraction of capacity and introduce skew when the \textsf{Premium:Freemium} ratio shifts.

\paragraph*{\textbf{Per-class latency at fixed load}}
\begin{figure*}[t]
    \centering
    \includegraphics[width=0.95\textwidth]{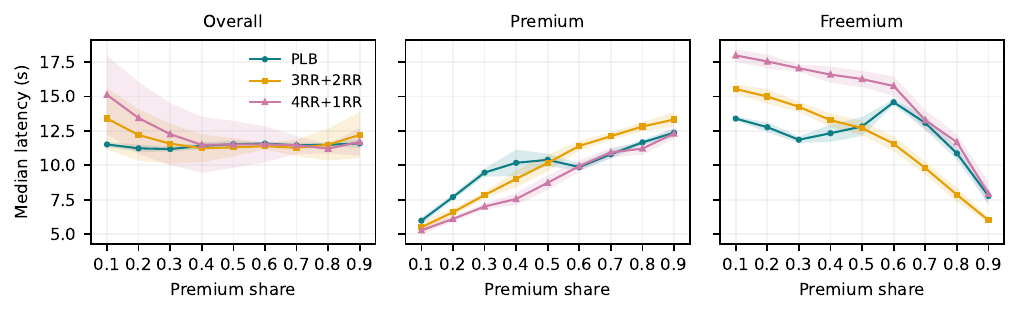}
    \vspace{-12pt}
    \caption{Per-class median latency at \(T = 1000\) terminals.}
    \label{fig:latency-triptych-1000}
\end{figure*}

Across mixes, \textsf{5PLB} keeps overall median latency in a narrow band, whereas each dedicated partition is competitive only near the \textsf{Premium} share for which it is implicitly tuned. 
As the \textsf{Premium} share moves away from those points, overall latency under \textsf{3RR+2RR} or \textsf{4RR+1RR} deteriorates substantially; in the worst Freemium-heavy case, one of the dedicated partitions is more than 30\% slower than \textsf{5PLB}. 
This mirrors the CPU results in Figure~\ref{fig:cpu-util-cv-1000}: \textsf{5PLB} uses the replica budget consistently across mixes, whereas static partitions incur penalties when their fixed split no longer matches the workload.

From the \textsf{Premium} perspective, dedicated partitions are most advantageous when \textsf{Premium} is a minority, since they reserve three or four replicas exclusively for a small number of \textsf{Premium} sessions. 
In these mixes, \textsf{3RR+2RR} achieves \textsf{Premium} medians about 8--17\% lower than \textsf{5PLB}, and \textsf{4RR+1RR} about 12--25\% lower. 
\textsf{PLB} deliberately does not chase this extra headroom: it keeps more replicas accessible to \textsf{Freemium}, which generates most of the queries in this regime, rather than reserving most of the cluster for a small set of \textsf{Premium} sessions. 
As Section~\ref{subsec:results-plb-vs-rr} shows, this sacrifice is only relative to ideal dedicated partitions: even when \textsf{Premium} traffic is modest, \textsf{Premium} latency under \textsf{5PLB} remains comparable to, or better than, the shared \textsf{5RR} baseline. 
Once \textsf{Premium} traffic becomes substantial, the situation reverses: \textsf{5PLB} closely tracks the best dedicated partition and is up to 16\% better than \textsf{3RR+2RR}.

\textsf{Freemium} exhibits the symmetric effect. 
When \textsf{Freemium} dominates, \textsf{5PLB} achieves the lowest \textsf{Freemium} median, reducing it by roughly 14--17\% compared to \textsf{3RR+2RR} and by about 25--30\% compared to \textsf{4RR+1RR}. 
Around a balanced mix, the \textsf{5PLB} \textsf{Freemium} median lies between the two dedicated partitions and remains close to the better one. 
As \textsf{Premium} becomes clearly dominant, \textsf{PLB} gradually withdraws borrowed capacity from \textsf{Freemium}: the best static partition can then make \textsf{Freemium} about 20--25\% faster than under \textsf{5PLB}, while the other remains within a few percent. 
This is the intended trade-off under heavy contention: \textsf{PLB} is willing to slow down the low-priority class relative to the most Freemium-friendly partition when \textsf{Premium} drives the load, to keep \textsf{Premium} latency low.

Together with the CPU results, these curves show that \textsf{5PLB} meets \hyperlink{obj:o3}{O3} while preserving the intended class-level trade-offs of \hyperlink{obj:o1}{O1} and \hyperlink{obj:o2}{O2}: overall latency stays within a few percent of the best dedicated configuration, \textsf{Premium} approaches its best achievable latency once it becomes the majority, and \textsf{Freemium} benefits when it dominates the workload.

\subsection{PLB vs. Round-Robin: Cost of Enforcing Priority}\label{subsec:results-plb-vs-rr}
We now compare \textsf{PLB} against the shared, priority-agnostic round-robin baseline \textsf{RR} to quantify both the \textsf{Premium} benefit of enforcing priority (\hyperlink{obj:o1}{O1}) and the corresponding cost for \textsf{Freemium} behavior (\hyperlink{obj:o2}{O2}).
We set the replica budget to 5 replicas and compare \textsf{5PLB} with \textsf{5RR} across 39 workload configurations, derived from 13 terminal counts and three representative \textsf{Premium:Freemium} mixes.

For each configuration, we collect per-run latency statistics for
\textsf{Premium} and \textsf{Freemium} and apply Welch's two-sample
\(t\)-test~\cite{welch1947} between \textsf{PLB} and \textsf{RR} on
both median and 95th-percentile latencies.

\begin{figure}[t]
    \centering
    \includegraphics[width=\columnwidth ]{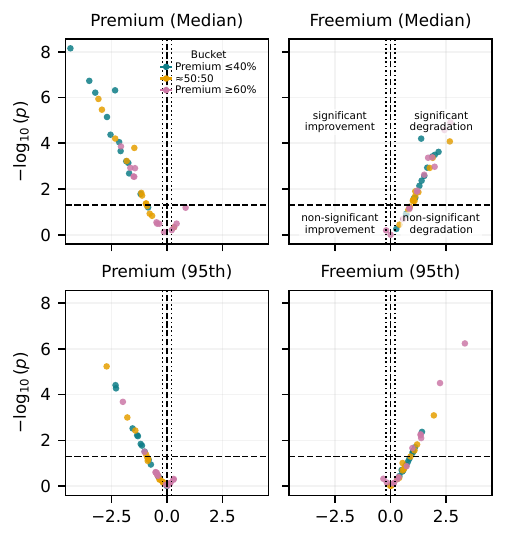}
    \vspace{-18pt}
    \caption{Latency volcano plots (\textsf{5PLB} vs \textsf{5RR}).}
    \label{fig:plb-vs-rr-volcano}
\end{figure}

We summarize the results using a volcano plot (Figure~\ref{fig:plb-vs-rr-volcano}), where each point corresponds to a single configuration.
The \(x\)-axis reports the standardized Hedges' effect size
\(g\)~\cite{hedges1981} between \textsf{PLB} and \textsf{RR}, and the
\(y\)-axis reports \(-\log_{10}(p)\), where \(p\) is the \(p\)-value of
the Welch test computed from per-run medians (top row) or
95th-percentile latencies (bottom row).
The negative values of \(g\) indicate lower latency under \textsf{PLB} than under \textsf{RR} (improvement), and positive values indicate higher latency under \textsf{PLB} (regression).
The horizontal dashed line marks the significance threshold
(\(p = 0.05\)), while the vertical dotted lines provide descriptive
reference bands for standardized effect sizes~\cite{cohen1988}.
We use these bands as visual guides rather than universal decision thresholds.

\paragraph*{\textbf{Premium latency}}
Across workloads, the \textsf{Premium} volcano plots are dominated by improvements. 
Out of 39 configurations, 26 show a statistically significant reduction in \textsf{Premium} median latency under \textsf{PLB}; the remaining configurations are not statistically distinguishable from \textsf{RR}. 
In those significant cases, \textsf{Premium} median latency typically improves by \(8\%\)–\(18\%\) (median \(11.5\%\)) relative to \textsf{RR}, and reaches up to \(28.4\%\) in the best case. At the 95th percentile, 14/39 configurations show significant \textsf{Premium} tail reductions of about \(10\%\), reaching up to \(14\%\), with no significant regressions. Once contention appears, \textsf{PLB} therefore shifts the \textsf{Premium} latency distribution towards lower values at both the median and the tail.

\paragraph*{\textbf{Freemium latency}}
For \textsf{Freemium}, the volcano plots show the expected cost of enforcing priority. 
Across 39 configurations, 28 show a significant increase in \textsf{Freemium} median latency under \textsf{PLB}, with an overhead of about \(11\%\) relative to \textsf{RR}. 
At the 95th percentile, 14 configurations show a significant slowdown of about \(7\%\), reaching up to \(15\%\) in the worst case. 
These slowdowns are concentrated in Premium-dominated mixes, where the policy is designed to protect \textsf{Premium}.
 
This shows that the cost of enforcing priority remains moderate: even when \textsf{Freemium} slows down under \textsf{PLB}, the degradation stays on the order of 10\% rather than factors, while overall latency remains close to \textsf{RR}, consistent with \hyperlink{obj:o2}{O2}.

Taken together, these results show that \textsf{PLB} meets \hyperlink{obj:o1}{O1} and \hyperlink{obj:o2}{O2}: it improves \textsf{Premium} latency over shared \textsf{RR} while keeping the induced \textsf{Freemium} slowdown controlled.

%% file: sections/related_work.tex
\section{Related work}

Prior work on load balancing and replica selection for distributed databases has primarily focused on optimizing performance, placement, or fairness without explicit consideration of user or service tiers. 
Techniques such as load balancing through replica swaps in multi-tenant databases~\cite{swat}, game-theoretic allocation in elastic database clusters~\cite{nashdb}, workload-aware replica selection~\cite{sword,tars}, and adaptive replica provision or fragment allocation under changing workloads~\cite{lion,fragalloc,dynloadbal} improve efficiency and responsiveness, but do not address service-tier differentiation at the session-routing layer. 
In contrast, \textsf{PLB} targets replicated databases under fixed resources and makes user priority an explicit signal in replica assignment.

Prior work has approached priority through a range of mechanisms, including admission control and query scheduling. 
SmartSLA, ActiveSLA, and Bouncer~\cite{smartsla,activesla,bouncer} decide whether or when to admit a request based on predicted system behavior and SLA objectives, while workload-aware schedulers and performance prediction methods prioritize concurrent queries inside the engine~\cite{selftuning,dbmspredict}. 
These approaches are complementary to ours, but operate at the level of admission or query execution rather than deciding, at session open, which replica should serve a given priority class.

More broadly, service differentiation and fixed-capacity resource management have been explored at the infrastructure level. 
Priority-aware resource sharing in edge and cloud systems~\cite{pars,edgematrix,slaaware,grsa} show the value of embedding service differentiation throughout the stack, while capacity-aware resource management approaches~\cite{smartsla,deralba} study efficiency under non-elastic resources. 
However, these works operate at the VM, application, or network layer, or do not incorporate service-tier awareness into replica assignment. 
\textsf{PLB} instead addresses priority-aware capacity sharing at the database session layer in replicated deployments.

%% file: sections/threats.tex
\section{Threats to Validity}

\paragraph*{Workload and system scope}
Our evaluation focuses on a read-only OLAP setting using TPC-H at a single scale factor on PostgreSQL~14 in a replicated deployment over homogeneous machines. 
This matches our target scenario, but does not cover OLTP or mixed read/write workloads, other DBMSs, replication schemes, or heterogeneous hardware.
In such settings, contention may arise differently, and we therefore do not expect the exact numerical results reported here to carry over unchanged. 

\paragraph*{Configuration and priority model}
\textsf{PLB} exposes a small set of parameters, which in this study are fixed once per replica budget using the calibration procedure of Section~\ref{sec:experimental-setup}; we do not explore the full parameter space or automatic tuning. We also restrict attention to two priority classes and session-level routing, assuming that sessions can be tagged with a class. 
This leaves richer service hierarchies and per-query priorities outside the scope of the present study.

%% file: sections/conclusion.tex
\section{Conclusion \& Perspectives}
This paper addressed priority-aware routing in replicated databases under a fixed replica budget. 
We presented \textsf{PLB}, a session routing strategy that partitions replicas by priority class and borrows capacity across classes under contention without modifying the database engine.

Our evaluation shows that \textsf{PLB} provides a middle ground between static dedicated partitions and a fully shared round-robin pool. 
Compared to dedicated clusters, it avoids underutilized replicas and large load imbalances when the \textsf{Premium:Freemium} mix changes, while keeping latency close to that of the best static allocation. 
Compared to round-robin, it improves \textsf{Premium} median latency by about 11--12\% on average, up to 28\%, while keeping the \textsf{Freemium} overhead around 10\% and leaving overall latency largely unchanged. 
Overall, \textsf{PLB} achieves prioritized latency for high-priority users, controlled degradation for low-priority users, and efficient sharing of a fixed replica budget.

Although we evaluate \textsf{PLB} on a replicated relational database, the same approach to priority-aware capacity sharing may also apply to other replicated, read-heavy services operating under fixed resource budgets. 
Likewise, the \textsf{Premium}/\textsf{Freemium} classes used in our experiments are only one interpretation of priority: the same mechanism could also differentiate applications, tenants, or request types with different latency requirements.

Future work includes extending \textsf{PLB} beyond two priority classes, exploring query-level routing, and studying transactional, mixed, and multi-tenant workloads.